\documentclass[letterpaper,twoside,twocolumn,english,aps,prl]{revtex4}
\usepackage{amsmath}
\usepackage{graphicx}
\usepackage{amssymb}

\begin{document}

\title{Spin relaxation and antisymmetric exchange in $n$-doped III-V semiconductors}

\author{L.P. Gor'kov and P.L. Krotkov}

\affiliation{National High Magnetic Field Laboratory, Florida State University,
Tallahassee FL 32310,}

\affiliation{L.D. Landau Institute for Theoretical Physics, Russian Academy of
Sciences, 2 Kosygina st., 117334 Moscow Russia}

\date{November 6, 2002}

\begin{abstract}
Recently K. Kavokin {[}Phys. Rev. \textbf{B 64}, 075305 (2001){]}
suggested that the Dzyaloshinskii-Moriya interaction between localized
electrons governs slow spin relaxation in $n$-doped GaAs in the regime
close to the metal-insulator transition. We derive the correct spin
Hamiltonian and apply it to the determination of spin dephasing time
using the method of moments expansion. We argue that the proposed
mechanism is insufficient to explain the observed values of the spin
relaxation time.

PACS number(s): 71.70.Gm, 71.70.Ej, 76.30.Da
\end{abstract}
\maketitle
\newcommand{\f}[1]{\mbox{\boldmath\(#1\)}}

\newcommand{\ff}[1]{\mbox{\scriptsize\boldmath\(#1\)}}

\newcommand{\ds}[1]{\displaystyle #1}

\newcommand{\ts}[1]{\textstyle #1}

\newcommand{\sgn}{\mathop{\mathrm{sgn}}\nolimits}

\newcommand{\Tr}{\mathop{\mathrm{Tr}}\nolimits}

\newcommand{\so}{\epsilon}

Observation of exceedingly long spin relaxation times for electrons
optically pumped into the conduction band of $n$-doped GaAs \cite{kikkawa98}
prompted suggestions for the use of polarized spins in quantum computing
\cite{Prinz98}. No consensus has been reached, however, on the mechanisms
governing the spin relaxation processes. Low-temperature spin dephasing
time $\tau _{s}$ in Ref. \cite{kikkawa98} behaves non-monotonically
as a function of donor concentration $n_{D}$, reaching values above
100 ns at $n_{D}=10^{16}$ cm$^{-3}$. This concentration still belongs
to the nonmetallic regime.

In an interesting paper \cite{kavokin01} the author introduced a
new dephasing mechanism acting between the spins of electrons localized
on randomly distributed shallow donors. According to Ref. \cite{kavokin01},
dephasing would occur each time the spins of the two electrons on
neighboring centers rotate against each other in an effective spin-orbital
field in the process of tunneling between the two centers during the
spin flip-flop act caused by the exchange interaction. This notion
and the way it was treated in Ref. \cite{kavokin01} in terms of the
exchange narrowing does not seem to be fully self-consistent. The
point is that ``random effective fields'' originating in \cite{kavokin01}
from the antisymmetric Dzyaloshinskii-Moriya (DM) \cite{dzyaloshinskii57,moriya60}
interaction and the dynamical averaging due to the isotropic exchange
in fact both arise from the interaction of the spins of the very same
pair and hence are not independent.

As a more consistent approach to the problem we tried its formulation
in terms of the EPR line shape for the spin Hamiltonian which for
semiconductors without an inversion center is a sum of the conventional
isotropic and the antisymmetric DM exchange terms. The latter is responsible
for non-conservation of the total spin of the system.

The antisymmetric interaction \begin{equation}
\hat{\mathcal{H}}_{\mathrm{DM}}=\mathbf{d}(\mathbf{R})(\hat{\mathbf{s}}_{1}\times \hat{\mathbf{s}}_{2})\label{eq:DM}\end{equation}
in semiconductors lacking the inversion symmetry (III-V and II-VI
compounds) is of interest in itself, but has not been properly calculated
yet. In Ref. \cite{kavokin01} the asymptotic behavior of $\mathbf{d}(\mathbf{R})$
at large distances between centers $R\gg a_{\mathrm{B}}$, where $a_{\mathrm{B}}$
is the effective Bohr radius, for shallow (hydrogen-like) donor centers
was calculated in the Heitler-London approximation. But it has been
shown \cite{gorkov64,herring64} that the Heitler-London theory does
not account for the Coulomb correlations inside the ``tunneling corridor''
connecting the two centers which dramatically affects the prefactor
of the isotropic exchange. 

Below we derive the asymptotic form of $\mathbf{d}(\mathbf{R})$ at
$R\gg a_{\mathrm{B}}$. We have slightly modified the approach of
Ref. \cite{gorkov64,herring64} to include the initial spin degeneracy
of the problem and its lifting by interactions in the presence of
a weak spin-orbital interaction \begin{equation}
\hat{\mathcal{H}}_{\mathrm{so}}=\mathbf{h}(\hat{\mathbf{p}})\hat{\mathbf{s}}.\label{eq:so0}\end{equation}

Our theoretical result for the $R$-dependence of the antisymmetric
exchange $\mathbf{d}(\mathbf{R})$ differs significantly from Ref.
\cite{kavokin01,dzhioev02,footnote}. In Ref. \cite{dzhioev02} the
authors concluded that their experimental results for the spin relaxation
time $\tau _{s}$ together with the data \cite{dzhioev97,kikkawa98}
agree well with the mechanism \cite{kavokin01} in the concentration
range $n_{D}=2\times 10^{15}\div 2\times 10^{16}$ cm$^{-3}$. Upon
substituting into the same formulae of \cite{kavokin01,dzhioev02},
our corrected DM term increases the estimate for $\tau _{s}$ by a
factor of 4. Although this does not rule out completely the relaxation
mechanism \cite{kavokin01} because of the phenomenological character
of the analysis of the experimental data in \cite{dzhioev02}, the
possibility that the antisymmetric exchange alone accounts for the
values of $\tau _{s}$ observed in \cite{kikkawa98,dzhioev97,dzhioev02}
in this doping interval seems rather unlikely.

The essence of the method is as follows. The envelope of the wavefunction
of an electron localized on a donor impurity in the effective mass
theory satisfies Schr\"odinger equation with the Hamiltonian\begin{equation}
\hat{\mathcal{H}}=\mathbf{p}^{2}/2m_{e}+\hat{\mathcal{H}}_{\mathrm{so}}+e^{2}/\kappa r,\label{eq:1SE}\end{equation}
where $m_{e}$ is the effective electron mass, $\kappa $ is the permittivity.
(For GaAs $m_{e}\approx 0.072m_{0}$ and $\kappa \approx 12.5$). 

The spin-orbital Hamiltonian (\ref{eq:so0}) in zinc-blende semiconductors
has the form \cite{dresselhaus55,DP}\begin{equation}
h_{x}(\mathbf{p})=\frac{\alpha _{\mathrm{so}}}{m_{e}\sqrt{2m_{e}E_{g}}}p_{x}(p_{y}^{2}-p_{z}^{2}).\label{eq:so}\end{equation}
 $h_{y}(\mathbf{p})$ and $h_{z}(\mathbf{p})$ are obtained by permutation
in (\ref{eq:so}). Here $E_{g}$ is the band gap and $\alpha _{\mathrm{so}}$
is a phenomenological parameter. For GaAs $E_{g}\approx 1.43$ eV
and $\alpha _{\mathrm{so}}\approx 0.07$. 

Without the spin-orbital interaction the ground state had the energy
$\frac{1}{2}E_{0}=-\mathrm{Ry}=-m_{e}e^{4}/2\kappa ^{2}$. ($E_{0}\approx -12.6$
meV for GaAs). The corresponding eigenfunction of (\ref{eq:1SE})
were $\varphi _{\mu }^{\alpha }(\mathbf{r})=\varphi _{0}(r)\delta _{\mu }^{\alpha }$,
where $\delta _{\mu }^{\alpha }\equiv \delta _{\uparrow ,\downarrow }^{\alpha }$
denotes the spinors for up and down spin projection, \begin{equation}
\varphi _{0}(r)=\frac{1}{\sqrt{\pi a_{\mathrm{B}}^{3}}}e^{-r/a_{\mathrm{B}}},\end{equation}
 and $a_{\mathrm{B}}=\kappa /m_{e}e^{2}$ is the Bohr radius ($a_{\mathrm{B}}\approx 92\textrm{Å}$
for GaAs). In the presence of a spin-orbital interaction the spin
projection no longer commutes with (\ref{eq:1SE}), but the two-fold
(Kramers) degeneracy of the ground state remains. For small spin-orbital
interaction (\ref{eq:so}) eigenfunctions of (\ref{eq:1SE}) are \begin{equation}
\varphi _{\mu }^{\alpha }(\mathbf{r})\approx \varphi _{0}(r)(e^{im_{e}\hbar \mathbf{h}(\hat{\mathbf{r}})r\ff \sigma /2a_{\mathrm{B}}^{2}})_{\alpha \mu },\label{eq:phase}\end{equation}
where $\hat{\mathbf{r}}=\mathbf{r}/r$. The phase factor turns into
unity when $r\to 0$. Thus the Kramers index $\mu $ takes on the
meaning of a spin projection of an electron on the center.

The two-electron Hamiltonian is obtained by combining two single-electron
Hamiltonians (\ref{eq:1SE}) and the Coulomb interaction of electrons
between themselves and with the alien ions: \begin{eqnarray}
\hat{\mathcal{H}}_{\mathrm{tot}} & = & \left(\mathbf{p}_{1}^{2}+\mathbf{p}_{2}^{2}\right)/2m_{e}+\hat{\mathcal{H}}_{\mathrm{so}1}+\hat{\mathcal{H}}_{\mathrm{so}2}\label{2SE}\\
 & + & \frac{e^{2}}{\kappa }\left(\frac{1}{r_{12}}+\frac{1}{R}-\frac{1}{r_{1\mathrm{A}}}-\frac{1}{r_{1\mathrm{B}}}-\frac{1}{r_{2\mathrm{A}}}-\frac{1}{r_{2\mathrm{B}}}\right).\nonumber 
\end{eqnarray}
Here $r_{1\mathrm{A},\mathrm{B}}$ and $r_{2\mathrm{A},\mathrm{B}}$
denote the distances between the electrons and the two donor centers
A and B.

For isolated donors, i.e. on large distances $R\gg a_{\mathrm{B}}$
between two centers a two-electron wavefunction $\psi _{\mu \nu }^{\alpha \beta }$
is merely a product $\varphi _{\mu }^{\alpha }(\mathbf{r}_{1\mathrm{A}})\varphi _{\nu }^{\beta }(\mathbf{r}_{2\mathrm{B}})$
of two one-electron wavefunctions (\ref{eq:phase}). For large but
finite $R$ these functions have to be corrected to account for potentials
of the Coulomb interactions of electrons between themselves and with
the alien ions (\ref{2SE}). Following \cite{gorkov64,herring64},
one starts with the Schr\"odinger equation for the corrected functions
$\psi _{\mu \nu }^{\alpha \beta }$ \begin{equation}
\hat{\mathcal{H}}_{\mathrm{tot}}^{\alpha \alpha ',\beta \beta '}\psi _{\mu \nu }^{\alpha '\beta '}=E_{0}\psi _{\mu \nu }^{\alpha \beta }.\label{eq:B}\end{equation}

Let $\hat{\f \zeta }$ denote the direction from the donor A to the
donor B, positioned at $\zeta =\mp R/2$ respectively. It turns out
\cite{gorkov64} that for the calculation of exchange interaction
it is enough to know $\psi _{\mu \nu }$ on the median hyperplane
$\zeta _{1}=\zeta _{2}$ far from the centers. There one may seek
for the functions $\psi _{\mu \nu }$ approximately in the form \begin{equation}
\psi _{\mu \nu }^{\alpha \beta }(\mathbf{r}_{1},\mathbf{r}_{2})=\chi ^{\alpha \alpha ',\beta \beta '}(\mathbf{r}_{1},\mathbf{r}_{2})\varphi _{\mu }^{\alpha '}(\mathbf{r}_{1\mathrm{A}})\varphi _{\nu }^{\beta '}(\mathbf{r}_{2\mathrm{B}}),\label{eq:AB}\end{equation}
where $\chi ^{\alpha \alpha ',\beta \beta '}$ varies on the scale
of the order of $R$. 

Substituting (\ref{eq:AB}) into (\ref{eq:B}) yields a differential
equation on $\chi ^{\alpha \alpha ',\beta \beta '}$. The boundary
condition is determined from the condition that when either $\mathbf{r}_{1\mathrm{A}}\to 0$
or $\mathbf{r}_{2\mathrm{B}}\to 0$ the function $\psi _{\mu \nu }^{\alpha \beta }$
should convert to $\varphi _{\mu }^{\alpha }(\mathbf{r}_{1\mathrm{A}})\varphi _{\nu }^{\beta }(\mathbf{r}_{2\mathrm{B}})$,
i.e. $\chi ^{\alpha \alpha ',\beta \beta '}\to \delta ^{\alpha \alpha '}\delta ^{\beta \beta '}$.
The principal terms in spin-orbital interaction are included into
definition (\ref{eq:phase}), and the equation for $\chi ^{\alpha \alpha ',\beta \beta '}$
turns out to coincide with the one in the absence of the spin-orbit
\cite{gorkov64}\begin{equation}
\left(\partial _{\zeta _{1}}-\partial _{\zeta _{2}}-\frac{2}{R-2\zeta _{1}}-\frac{2}{R+2\zeta _{2}}+\frac{1}{R}+\frac{1}{r_{12}}\right)\chi =0.\label{eq:10}\end{equation}
Here we used that $\partial \varphi \sim \varphi /a_{\mathrm{B}}$
and hence one can neglect all but the first derivatives of the functions
$\chi $ which varies on distances of order $R\gg a_{\mathrm{B}}$. 

Therefore $\chi ^{\alpha \alpha ',\beta \beta '}=\chi _{0}\delta ^{\alpha \alpha '}\delta ^{\beta \beta '}$,
where on the median 5-dimensional hyperplane $\zeta _{1}=\zeta _{2}\equiv \zeta $
\cite{gorkov64}\begin{equation}
\chi _{0}=\frac{2R}{R+2|\zeta |}\sqrt{\frac{2\rho _{12}}{R-2|\zeta |}}e^{\frac{2|\zeta |-R}{2R}}.\label{chi}\end{equation}
Here $\rho _{12}=|\f \rho _{1}-\f \rho _{2}|$, and $\f \rho _{1,2}$
are the transverse radii-vectors of the electrons. 

The two-electron eigenfunction $\psi ^{\alpha \beta }(\mathbf{r}_{1},\mathbf{r}_{2})$
of $\hat{\mathcal{H}}_{\mathrm{tot}}$ adjusted for the exchange is
then sought in the form\begin{equation}
\psi ^{\alpha \beta }=\sum _{\mu ,\nu }c_{\mu \nu }\psi _{\mu \nu }^{\alpha \beta }+\sum _{\mu ,\nu }c_{\nu \mu }^{(P)}\psi _{\nu \mu }^{\alpha \beta },\label{expan}\end{equation}
where $c_{\mu \nu }$ and $c_{\mu \nu }^{(P)}$ are unknown coefficients.
$\psi _{\nu \mu }^{\alpha \beta }$ stands for the interchanged states
of the two electrons, when the first electron is now almost localized
on atom B in the state $\nu $ and the second --- on atom A in the
state $\mu $\begin{equation}
\psi _{\nu \mu }^{\alpha \beta }(\mathbf{r}_{1},\mathbf{r}_{2})\equiv \psi _{\mu \nu }^{\beta \alpha }(\mathbf{r}_{2},\mathbf{r}_{1}).\end{equation}

Because the total fermionic wavefunction should be antisymmetric in
particle permutation $\psi ^{\alpha \beta }(\mathbf{r}_{1},\mathbf{r}_{2})=-\psi ^{\beta \alpha }(\mathbf{r}_{2},\mathbf{r}_{1})$,
the coefficients $c_{\mu \nu }^{(P)}=-c_{\mu \nu }$.

We can now proceed to the derivation of the total exchange Hamiltonian.
Multiplying the Schr\"odinger equation $E\psi =\hat{\mathcal{H}}\psi $
by $\psi _{\mu \nu }^{*}$ (for brevity we omit spinor structure),
and the Schr\"odinger equation (\ref{eq:B}) $\hat{\mathcal{H}}^{*}\psi _{\mu \nu }{}^{*}=E_{0}\psi _{\mu \nu }^{*}$
by $\psi $, subtracting term by term and integrating over the region
$\zeta _{1}<\zeta _{2}$, we find \begin{equation}
(E-E_{0})c_{\mu \nu }=\int _{\zeta _{1}<\zeta _{2}}\left(\psi _{\mu \nu }^{*}\hat{\mathcal{H}}\psi -\psi \hat{\mathcal{H}}^{*}\psi _{\mu \nu }^{*}\right)d^{6}\mathbf{r}_{1,2}.\label{eq:12}\end{equation}

The kinetic part of $\hat{\mathcal{H}}$ can be readily reduced to
a surface integral \cite{gorkov64,herring64}\begin{equation}
(E-E_{0})c_{\mu \nu }=\frac{1}{2}\int _{\zeta _{1}=\zeta _{2}}\left(\psi \f \partial \psi _{\mu \nu }^{*}-\psi _{\mu \nu }^{*}\f \partial \psi \right)\mathbf{dS}\label{eq:16}\end{equation}
over the median plane. Here $\mathbf{dS}=(\hat{\f \zeta }_{1},-\hat{\f \zeta }_{2})d\zeta d^{4}\f \rho _{1,2}$
is its surface element and $\f \partial =(\f \partial _{1},\f \partial _{2})$
is the six-dimensional gradient.

We then substitute (\ref{expan}) into (\ref{eq:16}) neglecting derivatives
of $\chi _{0}$ compared to those of $\varphi _{\mu }$: \begin{equation}
\partial _{\zeta _{1,2}}\psi _{\mu \nu }\approx \mp \psi _{\mu \nu },\qquad \partial _{\zeta _{1,2}}\psi _{\nu \mu }\approx \pm \psi _{\nu \mu }.\label{eq:dpsi}\end{equation}

After simple calculations Eq. (\ref{eq:16}) results in \begin{eqnarray}
(E-E_{0})c_{\mu \nu } & = & -2c_{\mu '\nu '}^{(P)}\int _{\zeta _{1}=\zeta _{2}}\psi _{\mu \nu }^{*}\psi _{\nu '\mu '}d\zeta d^{4}\f \rho _{1,2}\label{eq:17}\\
 & = & J(R)(e^{-i\ff \gamma \ff \sigma /2})_{\mu \nu '}(e^{i\ff \gamma \ff \sigma /2})_{\nu \mu '}c_{\mu '\nu '}^{(P)}\label{eq:}\nonumber 
\end{eqnarray}
with the angle\begin{equation}
\f \gamma =m_{e}\hbar \mathbf{h}(\hat{\f \zeta })R/a_{\mathrm{B}}^{2}.\label{eq:gamma}\end{equation}
The exchange integral equals \cite{gorkov64,herring64}\begin{eqnarray}
J(R) & = & -2\! \! \! \int \limits _{\zeta _{1}=\zeta _{2}}\! \! \! \chi _{0}^{2}\varphi _{0}^{*}(r_{1\mathrm{A}})\varphi _{0}^{*}(r_{2\mathrm{B}})\varphi _{0}(r_{2\mathrm{A}})\varphi _{0}(r_{1\mathrm{B}})d\zeta d^{4}\f \rho _{1,2}\label{eq:}\nonumber \\
 & \approx  & -0.818\times 2\mathrm{Ry}\left(R/a_{\mathrm{B}}\right)^{5/2}e^{-2R/a_{\mathrm{B}}}.\label{eq:J}
\end{eqnarray}

Using \begin{equation}
c_{\mu \nu }^{(P)}=-c_{\mu \nu }\equiv -\ts \frac{1}{2}(\f \sigma _{\mu \mu '}\f \sigma _{\nu \nu '}+\delta _{\mu \mu '}\delta _{\nu \nu '})c_{\nu '\mu '}.\label{eq:13}\end{equation}
 we finally arrive at the equation for the coefficients $c_{\mu \nu }$,
$(E-E_{0})c=\hat{\mathcal{H}}_{\mathrm{ex}}c$, with the Hamiltonian
\begin{eqnarray}
\hat{\mathcal{H}}_{\mathrm{ex}} & = & -\ts \frac{1}{2}Je^{-i\ff \gamma (\ff \sigma _{1}-\ff \sigma _{2})/2}\left(\hat{\f \sigma }_{1}\hat{\f \sigma }_{2}+1\right)\label{eq:ee1}\\
 & \equiv  & -\ts \frac{1}{2}J\left[\hat{\f \sigma }_{1}\overleftrightarrow{R}(\f \gamma )\hat{\f \sigma }_{2}+1\right]\nonumber \\
 & \approx  & -\ts \frac{1}{2}J\left[\hat{\f \sigma }_{1}\hat{\f \sigma }_{2}+1-\f \gamma \left(\hat{\f \sigma }_{1}\times \hat{\f \sigma }_{2}\right)\right],\nonumber 
\end{eqnarray}
where the Pauli matrices $\hat{\f \sigma }_{1}$ and $\hat{\f \sigma }_{2}$
are supposed to act in the space of the Kramers indices, and $\overleftrightarrow{R}(\f \gamma )$
is the three-dimensional rotation matrix on the angle $\gamma $ around
the axis $\hat{\f \gamma }$.

\begin{figure}
\includegraphics[  width=0.90\columnwidth]{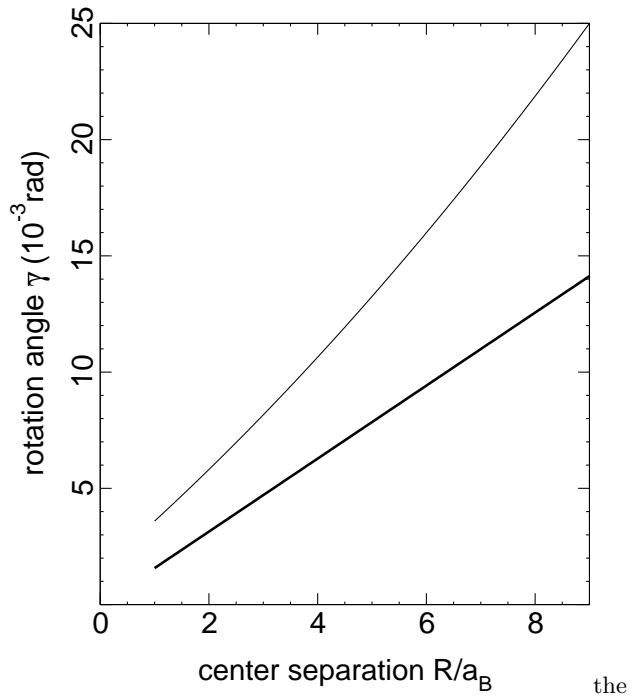}the

\caption{Mean square rotation angles $\overline{\gamma }$ of the asymmetric
interaction in GaAs as functions of the inter-center separations as
calculated in the present paper (thick line) and according to the
interpolation Eq. (7) in Ref. \cite{dzhioev02} (thin curve).\label{cap:Angles-of-rotation}}
\end{figure}

The rotation angle $\gamma $ (\ref{eq:gamma}) gives the strength
of the DM term relative to the isotropic exchange. It is perpendicular
to axis $\hat{\f \zeta }$ since $\mathbf{h}(\hat{\f \zeta })\hat{\f \zeta }=0$.
Taking the square mean of $\mathbf{h}(\hat{\f \zeta })$ over the
directions of $\hat{\f \zeta }$ we find that for GaAs\begin{equation}
\overline{\gamma }=\langle \gamma ^{2}\rangle _{\hat{\zeta }}^{1/2}=\frac{2\alpha _{\mathrm{so}}}{\sqrt{35E_{g}/Ry}}\frac{R}{a_{\mathrm{B}}}\approx 0.00157\frac{R}{a_{\mathrm{B}}}.\label{eq:28}\end{equation}
This $\overline{\gamma }$ is shown in Fig. \ref{cap:Angles-of-rotation}
against the $R$-dependence of $\overline{\gamma }$ from Ref. \cite[Eq. (7)]{dzhioev02}
for comparison. In the region of interest ($R\sim 3\div 7a_{\mathrm{B}}$)
our result for $\overline{\gamma }$ is about one half that of Ref.
\cite[Eq. (7)]{dzhioev02}.

We turn now to the question of how the antisymmetric exchange would
manifest itself as a spin relaxation mechanism. It was suggested in
Ref. \cite{kavokin01,dzhioev02} to write down the corresponding relaxation
time $\tau _{sa}$ in terms of a precession mechanism \cite{DP}:\begin{equation}
\tau _{sa}=\ts \frac{3}{2}\tau _{c}\overline{\gamma }^{-2},\label{eq:tsa}\end{equation}
where now, unlike \cite{DP}, $1/\tau _{c}$ is a rate of the flip-flop
transition caused by isotropic exchange. Inconsistency of the physical
picture of Ref. \cite{kavokin01} in the use of the dynamical averaging
in Eq. (\ref{eq:tsa}) was mentioned above. Another objectionable
point is to write $\tau _{c}$ in the form \cite{dzhioev02}\begin{equation}
\tau _{c}\approx \hbar /\xi J(R_{c}),\label{eq:tc}\end{equation}
where $R_{c}\approx \beta n_{D}^{-1/3}$, $\xi $ and $\beta $ being
the fitting parameters of the order of unity. The choice for the latter
one, $\beta \approx 0.65$, interpolates the maxima in the Poisson
distribution for the nearest and next nearest neighbors. Such a form
is assumed in Ref. \cite{dzhioev02} to phenomenologically account
for large average distances between donors. But since the maximum
in the Poisson distribution is rather broad, it appears erroneous
to merely substitute $R_{c}$ for $R$ in the exponential dependence
$J(R)$. 

At $n_{D}\approx 10^{16}$ cm$^{-3}$ the distance $R_{c}\approx 3.3a_{\mathrm{B}}$,
and from Fig. \ref{cap:Angles-of-rotation} for the angles $\overline{\gamma }$
\cite[Eq. (7)]{dzhioev02} one obtains $\overline{\gamma }(R_{c})\approx 0.009$,
while the corrected value in Eq. (\ref{eq:28}) $\overline{\gamma }\approx 0.005$.
The additional factor $\approx 4$ in Eq. (\ref{eq:tsa}) immediately
erodes the claimed quantitative agreement with experimental data in
Ref. \cite{kikkawa98,dzhioev02}.

With the applicability of (\ref{eq:tsa}) being questioned and the
experimental value of $\tau _{c}$ being not determined in \cite{dzhioev02}
unambiguously, we try to explore spin relaxation due to DM terms as
a question of the EPR line shape for a dilute system of paramagnetic
spins interacting via Hamiltonian (\ref{eq:ee1}). This appears a
proper approach if all other sources of inhomogeneous line broadening
are neglected and the limit is taken of the Larmor frequency $\omega _{\mathrm{L}}=g\mu _{\mathrm{B}}B\to 0$.

In Ref. \cite{larson89} such a problem dealing with DM interaction
has already been addressed for Mn-based II-VI-compound diluted magnetic
semiconductors in the framework of \emph{high temperature moment expansion}
of the linear response function\begin{equation}
\chi ''(\omega )=\frac{1}{2}\int _{-\infty }^{\infty }e^{-i\omega t}\langle [M_{x}(t),M_{x}(0)]\rangle dt.\label{eq:32}\end{equation}
 Here square brackets denote quantum commutator of the magnetization
operator $\mathbf{M}=g\mu _{\mathrm{B}}\sum _{n}\mathbf{s}_{n}$ (summation
runs over all donors). Angle brackets denote both thermodynamical
average and averaging over disorder.

The spin relaxation time is then given by \cite{larson89}\begin{equation}
\tau _{sa}=\sqrt{2I_{4}/\pi I_{2}^{3}},\end{equation}
where the frequency moments of the response function $I_{n}=\int _{-\infty }^{\infty }\omega ^{n-1}\chi ''(\omega )d\omega /\pi \chi (0)$
obtained in Ref. \cite{larson89} as a high-temperature expansion,
are\begin{eqnarray}
I_{2} & = & -\Tr ([\hat{\mathcal{H}}_{\mathrm{DM}},M_{x}]^{2})/\Tr (M_{x}^{2})+O(T^{-1}),\\
I_{4} & = & \Tr ([\hat{\mathcal{H}}_{\mathrm{ex}},[\hat{\mathcal{H}}_{\mathrm{DM}},M_{x}]]^{2})/\Tr (M_{x}^{2})+O(T^{-1}).
\end{eqnarray}

Calculating the commutators we find \begin{eqnarray}
I_{2} & = & 4n_{D}\int J^{2}(r)\left[\f \gamma ^{2}(\hat{\mathbf{r}})-\gamma _{x}^{2}(\hat{\mathbf{r}})\right]d^{3}\mathbf{r}+O(T^{-1}),\label{eq:I2}\\
I_{4} & \approx  & 32n_{D}\int J^{4}(r)\left[3\f \gamma ^{2}(\hat{\mathbf{r}})-\gamma _{x}^{2}(\hat{\mathbf{r}})\right]d^{3}\mathbf{r}+O(T^{-1}).\label{eq:I4}
\end{eqnarray}
(For an order of magnitude estimate below we neglected in (\ref{eq:I4})
the terms with double integration under the pretext that they are
of the ``second order'' in $n_{D}$. Averaging over $\hat{\mathbf{r}}$
we use $\langle \gamma _{i}(\hat{\mathbf{r}})\gamma _{j}(\hat{\mathbf{r}})\rangle _{\hat{r}}=\frac{1}{3}\overline{\gamma }^{2}\delta _{ij}$)

The high-temperature expansion (\ref{eq:I2}), (\ref{eq:I4}), where
one would integrate over all randomly distributed centers, applies
to weakly interacting spins only. Eqs. (\ref{eq:I2}), (\ref{eq:I4})
need to be corrected to take into account that if two spins are so
close that their exchange interaction $J(R)\gg T$, they would form
a singlet state and drop out of the system's thermodynamics. A proper
calculation of the temperature effects in the EPR line due to clusterization
should proceed in the framework of a scaling theory \cite{bhatt81}.
The problem is difficult and remains unsolved. We take the effects
into account qualitatively introducing a cut-off in the integrals
(\ref{eq:I2}), (\ref{eq:I4}) at short distances $a_{T}$ found from
the condition $J(a_{T})=T$. (At $T=2$ K this radius $a_{T}\approx 3.7a_{\mathrm{B}}$,
at $T=4.2$ K $a_{T}\approx 3.1a_{\mathrm{B}}$). 

Calculation of (\ref{eq:I2}), (\ref{eq:I4}) yields the answer in
terms of incomplete gamma functions $\Gamma (n,x)$:\begin{equation}
\tau _{sa}=\frac{6}{\pi n_{D}a_{\mathrm{B}}^{3}0.818\mathrm{Ry}\overline{\gamma }^{2}}\sqrt{\frac{2\Gamma (13,8a_{T}/a_{\mathrm{B}})}{\pi \Gamma ^{3}(8,4a_{T}/a_{\mathrm{B}})}}.\label{eq:myt}\end{equation}
For GaAs at $n_{D}=10^{16}$ cm$^{-3}$ at $T=2$K we find $\tau _{sa}\approx 290$
ns, and at $T=4.2$K we find $\tau _{sa}\approx 170$ ns as compared
to $\tau _{s}\approx 80$ ns found experimentally. (Using expression
(\ref{eq:tsa}) with corrected value for $\gamma $ from Fig. \ref{cap:Angles-of-rotation}
would give the temperature-independent value of $\tau _{sa}\approx 330$
ns at $n_{D}=10^{16}$ cm$^{-3}$) 

The method of taking a spin-orbital interaction into account, developed
in the present paper to shallow donor centers in \emph{bulk} zinc-blende
semiconductors, may be easily generalized. In a low-dimensional environment,
e.g., in quantum dots, spin-orbital interaction is described by a
linear dependence on $\mathbf{p}$: $h_{i}(\mathbf{p})=h_{ik}p_{i}$,
where $h_{ik}$ is a tensor. The spin structure of an exchange Hamiltonian
(\ref{eq:ee1}) acting on the coefficients of expansion (\ref{expan})
again decouples from the coordinate dependence $J(R)$. Although explicit
expression for $J(R)$, of course, now depends on the potential of
the quantum dots, the form of the ``rotated'' exchange Hamiltonian
(\ref{eq:ee1}) with the rotation angle $\gamma _{k}=m_{e}\hbar h_{ik}\hat{\zeta }_{k}R/a_{\mathrm{B}}^{2}$
instead of (\ref{eq:gamma}) remains. For a specific choice of the
linear in $\mathbf{p}$ spin-orbital interaction this result may be
obtained by an exact unitary transformation as has been recently shown
in \cite{Kavokin?}.

In the end, we comment on the role of the spin-orbital interaction
of electrons with the electric field of alien ions. For a homogeneous
field it has the form \cite{nozieres73} $\alpha _{E}\mu _{\mathrm{B}}g\hat{\mathbf{s}}(\hat{\mathbf{p}}\times \mathbf{E})2c/E_{g}$,
the coefficient $\alpha _{E}\sim 1$ depends on the band structure.
This term is proportional to $\sim \alpha _{E}(\mathrm{Ry}^{2}/E_{g})(m_{e}/m_{0})$,
and is much smaller than the term (\ref{eq:so}) considered above
that is proportional to $\sim \alpha _{\mathrm{so}}\mathrm{Ry}(\mathrm{Ry}/E_{g})^{1/2}$.

To summarize, we derived the asymptotically correct form of the spin
Hamiltonian for two hydrogen-like donors in the $n$-doped GaAs. We
applied the EPR line shape formalism to analyze spin dephasing times
$\tau _{sa}$ due to the antisymmetric DM exchange. The rough estimate
for $\tau _{sa}$ (neglecting the detailed low-temperature spin clusterization
which would further increase $\tau _{sa}$) yields values exceeding
those observed experimentally. Although the concentration range $n_{D}^{-1/3}\sim a_{\mathrm{B}}$
is very difficult for theory, our results suggest that the anisotropic
exchange between the localization centers is not a prevailing mechanism
of the spin relaxation in this concentration range.

The authors are thankful to V. Korenev and K. Kavokin for discussions
and sharing the details of interpretation of their data in \cite{dzhioev02}.
LPG thanks V. Dobrosavljevic, who attracted his attention to Ref.
\cite{bhatt81} and gratefully acknowledges fruitful conversations
on the subject with V. Kresin and E. Rashba. The work was supported
(LPG) by the NHMFL through the NSF cooperative agreement DMR-9527035
and the State of Florida, and (PLK) by DARPA through the Naval Research
Laboratory Grant No. N00173-00-1-6005.

\end{document}